# Langmuir-Blodgett Films a unique tool for molecular electronics

**Syed Arshad Hussain**

*Department of Physics, Tripura University, Suryamaninagar-799130, Tripura, India*
*Email: sa_h153@hotmail.com*

**Abstract:**

Molecular electronics is a new, exciting and interdisciplinary field of research. The main concern of the subject is to exploit the organic materials in electronic and optoelectronic devices. On the other hand Langmuir-Blodgett (LB) film deposition technique is one of the best among few methods used to manipulate materials in molecular level. In this article LB film preparation technique has been discussed briefly with an emphasize of its application towards molecular electronics.

## 1. Introduction

The past 30 years has witnessed the emergence of molecular electronics as an important technology for the 21st century [1-2]. Modern electronics is based largely on the inorganic semiconductor. In contrast an increasing number of organic materials are now finding use in the electronics industrial sector. The subject can broadly be divided into two main themes (although there is substantial overlap), as illustrated in figure 1.

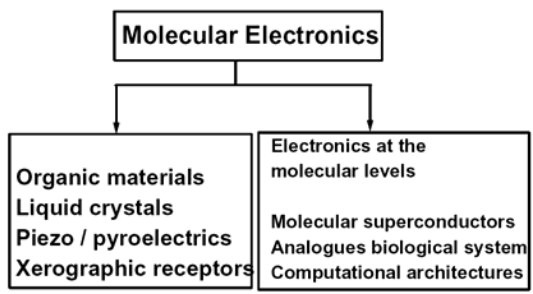

Figure-1: The sub-division of molecular electronics

The first concerns about the development of electronics and optoelectronic devices using the unique macroscopic properties of organics. This class of molecular electronics is already with us, the best example is liquid crystal display. Other areas in which organic compounds are becoming increasingly important are xerography, acoustic transducer (microphones and sonar devices) based on piezoelectric effect and pyroelectric sensors for infrared imaging.

The second strand to molecular electronics recognizes the dramatic size reduction in the individual processing elements in integrated circuits of recent years, as shown in figure 2.

In the last decade, the number of transistors on a silicon chip has increased by a factor 10[8]; the feature size on a current chip is now less than 1 $\mu$m [3]. The dimensions are comparable to those of large biological units. Therefore, molecular electronics deals with the manipulation of organic materials at the nanometer level to realize devices that will store and/or process information [4]. The exploitation of these nanostructures in instrumentation systems is still a long way off. Ideas in the area are limitless and the subject can all too easily merge with science fiction. However, there have been some determined research efforts.

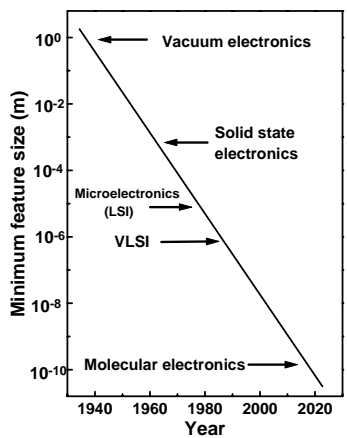

Figure 2: Electronic device sizes.

Certain organic compounds can possess semiconductive and even metallic properties [1]. It is therefore possible to fabricate diodes and field effect transistors (FET) using these compounds [5]. Organic light emitting displays may be based on a fluorescent organic material sandwiched between electrodes of high and low work functions [6]. Work on such devices is an emerging area from both academic and commercial point of view. However, over the next 20 years it is unlikely that organic materials will displace silicon as the dominant materials for fast signal processing. Silicon technology is firmly established and Moore's law, namely that device dimensions halved every two or three years, will probably describe the developments at least over the next decade. The semiconductor Industry Association bases a microelectronics roadmap on the future of CMOS technology. Its prediction is for silicon-based devices with a 35 nm minimum feature size and $10^8$ transistors per $cm^2$ by the year 2012.

The Langmuir-Blodgett (LB) technique provides one of the few methods of preparing organized molecular assemblies which are the pre-requisite for molecular electronic devices. LB film methods are perhaps the earliest example of what is now called 'supramolecular assembly', providing the opportunity to exercise molecular level control over the structure of organic thin films. It is the LB technique which trigger us to dream about the 'Molecular electronics' in which organic molecules perform an active function in the processing of information and in transmission and storage [7]. In this method, a single layer of molecules is first organized on a liquid surface, usually water, before being transferred onto a solid support to form a thin film with the thickness of a constituent molecule. If the process is repeated, multi-layered films can be prepared. The layer of molecules on a liquid surface is termed a Langmuir monolayer and after transfer it is called a Langmuir–Blodgett film. LB films provide a level of control over the orientation and placement of molecules in monolayer and multilayer assemblies that are difficult to otherwise achieve. LB films have been explored for applications that include electronics, optics, microlithography, and chemical sensors, as well as biosensors or biochemical probes [8-9]. Following sections of this article describes a brief introduction about the LB technique followed by few examples of using LB method to realize molecular electronic devices.

## 2. Historical overview

Benjamin Franklin is normally credited first for the discovery of floating monolayer thin films. In 1774 he observed the behavior of a small amount of oil in a pond. He saw that the oil calmed water ripples, induced by the outside wind, even when the oil film was so thin it was no longer visible. He furthermore noticed that the spreading of the oil on the water surface was very different from spreading on a glass surface, and he therefore questioned the reasons for all this different behavior.

The matter was however not studied further until Lord Rayleigh investigated the surface tension of oil films spreading on water. He managed to estimate that the thickness of those films was in the range 1-2 nm, which is close to the currently known thickness of monolayer oil films.

An amateur scientist, Agnes Pockels, described her design of a water basin with a movable surface barrier in a letter to Lord Rayleigh in 1891. In this way she paved the way for further monolayer research. This helped Rayleigh to measure thicknesses for other types of oil films, while Pockels herself studied the surface tension of different kind of oils.

Floating monolayer films are however named after the infamous scientist Irving Langmuir, researcher at the General Electric Company in the first half of the twentieth century, for his development of the experimental apparatus and his studies on the properties of a wide range of film materials in the 1910's and 1920's. Not only did he accurately determine the size of the molecules, but also described their orientation and structure.

The technique is named after Irving Langmuir and his research assistant Katharine Blodgett. Langmuir awarded the Nobel Prize in Chemistry in 1932 for his studies of surface chemistry, used floating monolayers to learn about the nature of intermolecular forces. In the course of his studies, Langmuir developed several new techniques that are by and large still used today in the study of monomolecular films. Together with Langmuir, Katharine Blodgett refined the method of transferring the floating monolayer onto solid supports. Although the methods are commonly named after Langmuir and Blodgett, numerous observations and experiments on floating organic films predate them, and enjoyable accounts of this history are available [7].

It was however not until 1965 that Langmuir-Blodgett films received much further attention, when Hans Kuhn studied their spectroscopic properties. His research seemed to initiate increased interest in the properties of Langmuir-Blodgett films, as since then the number of research groups has been increasing steadily, until perhaps in the recent years when the commercial implementation of LB films is getting closer to reality [10].

### 3. What makes LB films appealing?

The appealing feature of Langmuir-Blodgett films is the intrinsic control of the internal layer structure down to the molecular level and the precise control over the resulting film thickness. Sophisticated LB troughs allow us to process several materials with different functionalities and offer the possibility to tune the layer architecture according to the demands of the desired molecularly engineered organic thin-film devices [7-8, 11].

### 4. Basic concepts of Langmuir-Blodgett Films

#### 4.1. LB compatible materials

In order to form a Langmuir monolayer, it is necessary for a substance to be water insoluble and soluble in a volatile solvent like chloroform or benzene. LB compatible materials consist of two fundamental parts, a 'head' and a 'tail' part. The 'head' part is a hydrophilic (water loving) chemical group, typically with a strong dipole moment and capable of hydrogen bonding, like $-OH$, $-COOH$, $-NH_2$ etc. The 'tail' part on the other hand is hydrophobic (water repeling), typically consisting of a long aliphatic chain. Such molecules, containing spatially separated hydrophilic and hydrophobic regions, are called amphiphiles [10]. Typical example of LB compatible materials is the long chain fatty acids (stearic acid, arachhidic acid etc) and their salts. A schematic of LB compatible molecule is shown in figure 3.

If amphiphile molecules arrive at the air-water interface with its hydrophobic tails pointing towards the air and its hydrophilic group towards water, the initial high energy interface is replaced by lower energy hydrophilic – hydrophilic and hydrophobic – hydrophobic interfaces, thus lowering the total energy of the system. Hence, the molecules at the interface are anchored, strongly oriented normal to the surface and with no tendency to form a layer more than one molecule thick.

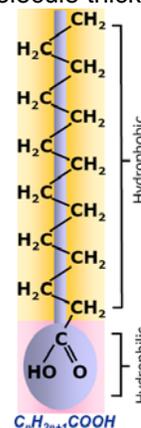

Figure 3: The general chemical structure of a LB compatible molecule with a carboxylic acid head and an arbitrary tail.

#### 4.2. Langmuir monolayer formation

Essentially all LB film works begin with the Langmuir-Blodgett trough, or Langmuir film balance, containing an aqueous subphase (figure 4). Moveable barriers that can skim the surface of the subphase permit the control of the surface area available to the floating monolayer. Now a days sophisticated Langmuir-Blodgett (LB) film deposition instrument are designed and marketed by several companies. A typical LB film deposition instrument installed in our laboratory (designed by Apex Instruments Co., India) is shown in figure 5.

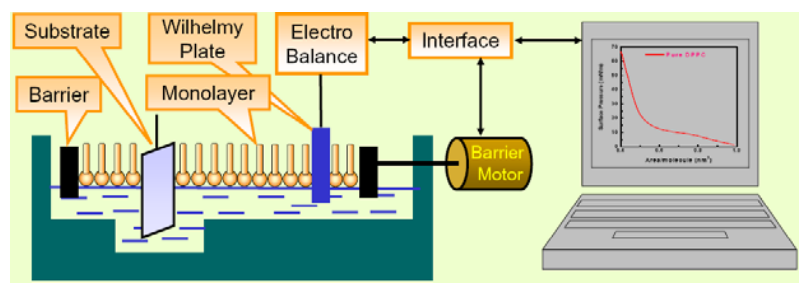

Figure 4: Schematic of LB trough. The Wilhelmy plate monitors the surface through a microbalance interfaced with computer. Barrier movement is also controlled by computer.

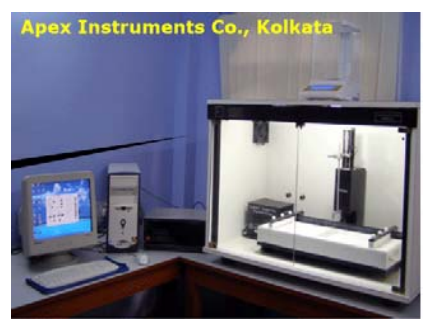

Figure 5: A typical LB film deposition instrument installed in our laboratory.

To form a Langmuir monolayer film, the molecules of interest is dissolved in a volatile organic solvents (chloroform, hexane, toluene etc.) that will not dissolve or react with the subphase. The dilute solution is then minutely placed on the subphase of the LB trough with a microliter syringe. The solvents evaporate quickly and the surfactant molecules spread over the subphase surface in the LB trough.

In order to control and monitor the surface pressure, $\pi$ (this quantity is the reduction of surface tension below that of clean water), the barrier intercepting the air-water interface is allowed to move so as to compress or expand the surface film. Wilhelmy plate arrangement is used to measure the surface pressure. In this method a small piece of hydrophilic material, usually filter paper, intercepting the air-water interface and supported from the arm of an electronic microbalance which is interfaced with a computer. The force exerted is directly proportional to the surface tension. There are several techniques available to monitor the state of the floating monolayer [7-8]

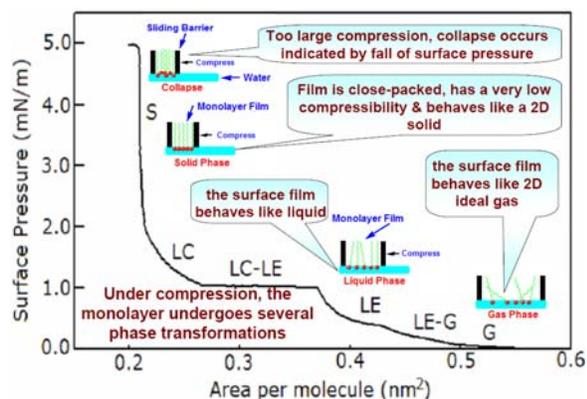

Figure 6: Schematic of surface pressure – area per molecule ($\pi$ -A) isotherm showing different phases of monolayer at air-water interface.

The measurement of surface pressure ($\pi$) as a function of area per molecule (A) in the monolayer films is known as the isotherm characteristics. This characteristic is easily obtained and much useful information about the mono-molecular films at the air-water interface. A conceptual illustration of the surface pressure versus area per molecule isotherm is shown in figure. As the pressure increases, the two-dimensional monolayer goes through different phases that have some analogy with the three-dimensional gas, liquid, and solid states. If the area per molecule is sufficiently high, then the floating film will be in a two-dimensional gas phase where the surfactant molecules are not interacting. As the monolayer is compressed, the pressure rises signaling a change in phase to a two-dimensional liquid expanded (LE) state, which is analogous to a three dimensional liquid. Upon further compression, the pressure begins to rise more steeply as the liquid expanded phase gives way to a condensed phase, or a series of condensed phases. This transition, analogous to a liquid–solid transition in three dimensions, does not always result in a true two-dimensional solid. Rather, condensed phases tend to have short-range structural coherence and are called liquid condensed (LC) phases. If the surface pressure increases much further the monolayer will ultimately collapse or buckle, not still being a single molecule in thickness everywhere. This is represented by a sudden dip in the surface pressure as the containment area is decreased further, such as is shown in figure 6.

### 4.3. Langmuir-Blodgett films

The term 'Langmuir–Blodgett film' traditionally refers to monolayers that have been transferred off of the water subphase and onto a solid support. The substrate can be made of almost anything. However the most common choices are as glass, silicon, mica or quartz etc. Vertical deposition is the most common method of LB transfer; however, horizontal lifting of Langmuir monolayers onto solid supports, called Langmuir–Schaeffer deposition, is also possible [7-8, 11].

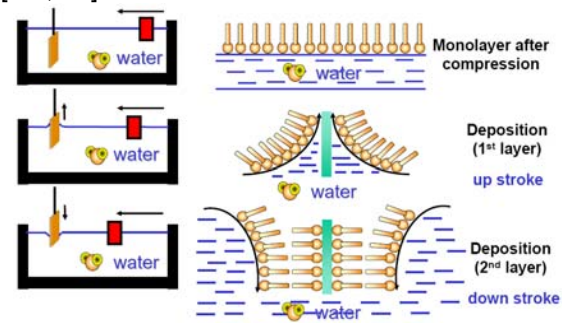

Figure 7: Deposition scheme during up and down stroke

In principle the Langmuir-Blodgett deposition method simply consists of dipping and pulling a solid substrate, orientated vertically, through the coating monolayer while keeping the surface pressure constant at a desired value (figure 7). It is also a common practice to coat the substrate with a highly hydrophobic or hydrophilic material. The rate at which the substrate is dipped or pulled through the monolayer must also be precisely controlled and kept constant at a very low value (typically 1 – 5 mm/min). The surface pressure for film deposition is normally chosen to be in the solid like region. However, at any pressure film can be deposited. The transfer of monolayer film occurs via hydrophobic interactions between the alkyl chains and the substrate surface or the hydrophilic interaction between the head groups of the molecules and the hydrophilic substrate surface. Subsequent dipping or pulling deposits a second layer on top of the first, the process simply being repeated until the desired number of layers has been deposited.

## 5. Research and applications

Modern studies of floating monolayers and LB films fall largely into two areas. The first area includes detailed fundamental studies of the physical nature and structure of Langmuir monolayers and LB films. The other involves applications that take advantage of the ability to prepare thin films with controlled thickness and composition. Much of the current work toward applications derives inspiration from the pioneering work of Hans Kuhn [12], who, in the 1960s, moved away from studies on fatty acids and other simple amphiphiles and used LB methods to control the position and orientation of functional molecules within complex assemblies. Few examples of using LB films in molecular electronics have been given in the following sections of this article.

### 5.1. Directional electron transfer

The elaborate molecular machinery of the photosynthetic pathway continues to inspire efforts to artificially control the alignment and orientation of molecular components to achieve complex functionality such as unidirectional electron transport. Langmuir–Blodgett deposition has been an important tool for preparing highly organized molecular assemblies in which intermolecular interactions such as distance, orientation, and extent of interaction can be controlled. In general, two different strategies have been employed to achieve vectorial electron transfer in LB films for applications that include photoinduced electron transfer and molecular rectifiers. The first approach takes advantage of the one-layer-at-a-time deposition process to position the various molecular components at different positions within a cooperative multilayer assembly. The second strategy is to use LB deposition to orient prefabricated molecular dyads or triads. The potential of using monolayer techniques to organize molecular components for electron transfer and energy transfer was demonstrated by Kuhn and co-workers [12] working with long-chain substituted cyanine or azo dyes mixed into fatty acid monolayers.

### 5.2. Organic conductors

Molecule-based conductors depend on intermolecular interactions and therefore the arrangement of molecules in condensed phases. Electronic band structures are determined by the distances between molecules and their orientation relative to each other. Despite significant advances in understanding the factors that dictate molecular packing, the ability to design a functional molecule and predict how it will arrange in the solid state is quite limited. The LB method provides at least some level of control over the orientation and placement of molecules in monolayer and multilayer assemblies [13].

### 5.3. Organic diodes

Since the discovery of semiconducting behaviour in organic materials, there has been a considerable research effort aimed at exploiting these properties in electronic and optoelectronic devices. Semiconducting LB films have been used to inorganic semiconductors (eg. Si, GaAs) in metal / semiconductor / metal structures. Perhaps the simplest example is that of a diode. Here, the LB film is sandwiched between metals of different work functions. In the ideal case, an n-type semiconductor should make an ohmic contact to a low work function metal and a rectifying Schottky barrier to a high work function metal [14]. The reverse is true for a p-type semiconductor. Figure 8 show the electronic potential energy band structure for a Schottky barrier formed by sandwiching pthalocyanine LB film between aluminium and indium-tin-oxide electrodes. The LB material is a p-type semiconductor and the aluminium / LB film interface provides the rectifying junction.

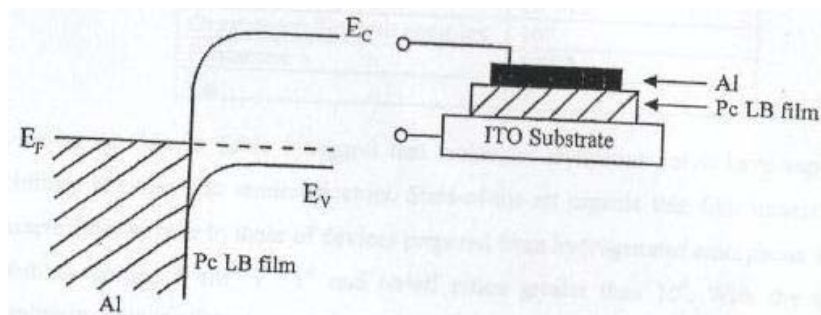

Figure 8: Energy band structure of a rectifying contact between an unsubstituted pthalocyanine LB films and an aluminium contact. The electrical contact between the indium-tin-oxide (ITO) substrate and the LB film is ohmic [14].

### 5.4. Organic field effect transistors (FET)

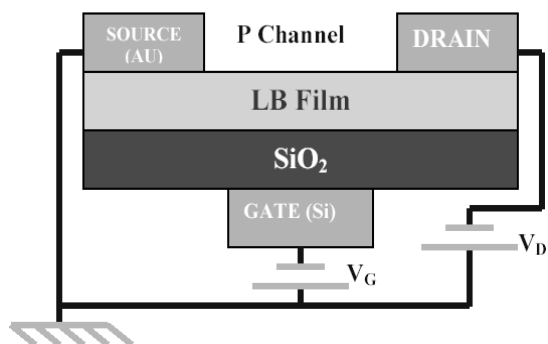

Figure 9: Schematic of OFET of pthalocyanine (PC) LB films [15]

Microelectronics technology is based on the ways of fabricating and manipulating thin layers. LB technique finds another interesting application in the fabrication of thin films of copper phthalocyanine derivatives as field effect transistors (FET). It is well known that pthalocyanine derivatives are very promising organic semiconductor materials due to their chemical and thermal stability. Among phthalocyanine, copper phthalocyanine derivatives have been utilized as organic field effect transistors (OFETs). The FET performances of the LB films of phthalocyanine has been tested by I-V curves acquired from devices operating in accumulation mode. It has been observed that to improve the carrier mobility of PC films, the arrangement through a more highly ordered film with improved interaction distance and $\pi - \pi$ interaction and decreases the effects to result in better quality of the LB films throughout the FET channel.

### 5.5. Optical Applications

Electroluminescence is the radiation of light from a material with an electric field applied across it.

This effect has been observed in some Langmuir-Blodgett films, which is convenient because of the low voltage requirement resulting from the extremely small thickness. For example, to achieve a specific light intensity only 6V are required for a LB films of certain material as compared to 200V for an evaporated film of the same material [16]. LB films have successfully been used to create a polymer light emitting diode (PLED), for example with the film being composed of poly fluorene monolayers [17]. However, commercial implementation, e.g. in the form of organic light emitting diodes (OLED), has likely not become a reality yet but that is probably only a question of time because of the low power consumption potential resulting from the use of very thin films. The use of LB films has also been suggested in order to create a photo-responsive conductivity switch. A working example of such a film is shown in figure 10. The film is in principle composed of two parts, a conductive part and a photo-responsive part. Exposure to a UV-light changes the photo-responsive part structurally, and since the conduction of the conductive part depends on the structure of the upper photo-responsive part the conduction of the film is increased. An exposure to a visible light then repairs the structure of the photo-responsive part to its original form, which also changes the conduction back to its low value [18].

The so called optical second harmonic generation (SHG) effect in some LB films has received considerable attention. This optical effect in principle halves the wavelength of a light incident on the film (or equivalently doubles the frequency of the light) and is caused by a nonlinear oscillation of the electric dipoles that are present in certain films. In those films, the absorption of light causes the oscillation of those dipoles at double the fundamental frequency, i.e. at the second harmonic frequency. This could for example be used to change the color of the invisible infrared YAG laser ($\lambda$ = 1.06 $\mu$m) to a visible green light ($\lambda$ = 532 nm). Electric dipoles (molecular level) can be build up in a LB film with either the X- or Z-type deposition scheme, if not depositing with the alternate layer method. Although theory predicts that the light intensity should increase quadratically with the film thickness, experiments have shown that the effect can only be observed if the film is relatively thin. Thus, LB films are generally considered a convenient way of reproducing this second harmonic generation effect. For this purpose, films made of for example [19] hemicyanine dyes (with the general chemical structure [D - $C_6H_4$ - CH = CH - $C_5N^+$ - RX⁻], where D is some electron donor, R is an alkyl chain and X is some negative ion) have been studied thoroughly [10].

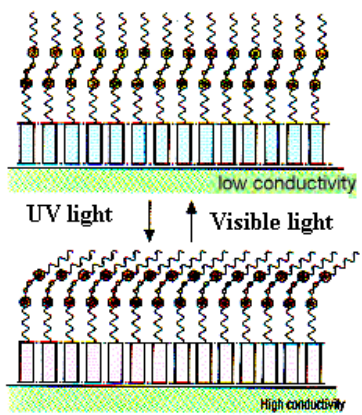

Figure 10: A schematic of a working example of a photo-responsive conductivity switch

Photoconductivity is the increase of the electrical conductivity of the material by the absorption of suitable radiation. It is widely used in various electronics products for example auto brightness control (ABC) circuits in TV sets, camera shutter, car dimmers, street light control, auto gain control in transceivers electrophotography etc. Polypyrrole Langmuir-Blodgett (LB) films (band gap 3.2 eV) produced electrochemically can be anticipated to exhibit good photoconductivity [20].

### 5.6. Pyroelectric effect

A material is said to be pyroelectric if it has a temperature dependent spontaneous polarization, i.e. the polarity of film's current is positive when the temperature is increasing and negative when it is decreasing. The material thus generates electric potential when heated or cooled and may therefore be used to detect temperature changes in the film's environment. The pyroelectric effect of a variety of Langmuir-Blodgett films has been studied quite well, resulting in a steady and quite significant improvement in the pyroelectric coefficient (the sensitivity of material to rate of temperature variations). [21].

### 5.7. LB Films as Sensors

A sensor is a physical device or biological organ that detects, or senses, a signal or physical condition and chemical compounds. It is basically consisting of a transducer and a selective sensing layer [22]. It is invariably provided by a material in which some selective interaction of the species of interest takes place which results in the change of some physical parameters such as electric current or potential or conductivity, intensity of light, mass, temperature etc.

Polypyrrol (Ppy) and its derivatives are widely used for the prepartaion of various types of sensors depending on their transducing mechanism. Ppy films deposited by LB technique were used selectively to detect ammonia [23].

### 5.8. LB films as Rectifier

Rectifying devices are the basic components, widely used in many electronic circuits. Polymer p-n junction and Schottky junction prepared by using the Langmuir-Blodgett (LB) technique have been studied extensively in recent years [24]. The n-type Semiconducting property of an anion doped polypyrrole/polythiopene has been demonstrated [25].

Electrical rectifying devices such as Zener diode have also been prepared using a two-layer configuration, consisting of a p-doped semiconducting polymer polypyrrole or poly (3-methylthiphene) layer and an n-type multilayer structure of CdSe and 1, 6-hexanedithiol [26].

### 5.9. LB Films as MIS Structure

Metal-Insulator-Semiconductor (MIS) structures were fabricated by vacuum deposition of various metals such as indium, aluminium and tin on LB films of cadmium stearate obtained on polypyrrole films deposited on indium tin oxide glass [27]. The value of the dielectric constant of the insulating $CdSt_2$ LB films was found to be 1.84, which is in good agreement with the experimental results reported earlier [28].

The alternating deposition of mono- or multilayer of undoped poly (3-hexylthiophene) and doped polypyrrole prepared by Langmuir-Blodgett (LB) technique have been used to fabricate the organic heterostructures [28].

### 5.10. LB Films in SMD

Single Molecule Detection (SMD) using Surface-Enhanced Resonance Raman Scattering and Langmuir-Blodgett (LB) films have also been investigated [29].

In this particular work, the Langmuir-Blodgett (LB) method has been employed in order to obtain the surface-enhanced resonance Raman scattering (SERRS) spectra of a single dye molecules in the matrix of a long chain fatty acid.

### 5.11. LB Films as Acoustics Surface Wave Devices

It is the natural orientation features of the Langmuir monolayers and also the degree of control over the molecular architecture which led the LB films in various applications such as acoustic surface wave devices, infra-red detectors and optoelectronics, where the materials with non-centrosymmetric structures are required [30].

### 5.12. Films with mixed properties

The layer-by-layer deposition process and the amphiphilic nature of the basis molecules make LB films ideal platforms for combining more than one targeted property into a single material. The film-forming molecules can be changed from one layer to the next to give superstructures in which the chemically different layers contribute different physical properties. Alternatively, the segregated hydrophobic and hydrophilic character of LB films can lead to mixed organic/inorganic films, or 'dual-network' assemblies, where the organic and inorganic components contribute separate properties [31-32]. An example is a film that is both conducting and magnetic. Films with photoactive or electroactive components designed to switch the property of interest with external stimulus have also been targeted.

The heterostructured character of hybrid LB films provides opportunity to explore the photochemical switching of magnetic behaviour by coupling a photoactive chromophore with a magnetic lattice.

### 6. Summary

Molecular electronics has given the scope to revolutionize material science, electronic and opto-electronic device research and ultimately to have a significant impact on instrumentation and measurement science. Langmuir-Blodgett (LB) film methods continue to provide routes to unique organic thin film materials. The key to LB techniques is the ability to control the organization of molecular component on a molecular level. The LB technique allows elegant experiments to be undertaken in the research laboratory that can provide a valuable insight into the physical processes that underpin the device operation. These works will also pave the way for the development of molecular scale electronic devices, which emulate natural processes. Therefore, development of LB films for practical applications is a challenge, requiring an interdisciplinary outlook which neither balks at the physics involved in understanding assemblies of partially disordered and highly anisotropic molecules nor at the cookery involved in making them. Although, LB films cannot be adapted to all purposes, there are signs that with sufficient understanding, their behaviour can be optimized for specific technological applications. In spite of several difficulties, LB films have a unique potential for controlling the structure of organized matter on the ultimate scale of miniaturization, and must surely find a niche where this potential is fulfilled.

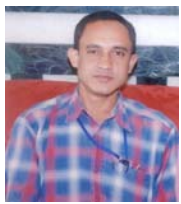

Dr. Syed Arshad Hussain is presently working as Assistant Professor in Physics in the Department of Physics, Tripura University, Tripura, India. His research interest includes Langmuir-Blodgett films, Self Assembled Films, Nano-Clay, biomolecules etc. He worked as Visiting Fellow of K. U. Leuven, Belgium from July, 2007 to August, 2008.
http://www.geocities.com/sa_h153; http://www.arshad.hdfree.in (personal); http://www.tripurauniversity.in (Office)
Email: sa_h153@hotmail.com